\documentclass[preprint,12pt]{elsarticle}

\usepackage{multicol}
\usepackage{multirow}
\usepackage{tabularx}
\usepackage{color,soul}

\usepackage{comment}
\usepackage{natbib}
\usepackage{mathtools}
\usepackage{caption}
\usepackage{subcaption}
\usepackage{bibunits}
\usepackage[colorinlistoftodos]{todonotes}
\usepackage{framed}
\usepackage[colorlinks=true, allcolors=blue]{hyperref}
\usepackage{nomencl}    

\usepackage{lineno}

\sethlcolor{yellow}
\usepackage{setspace}

\usepackage{url,lineno,microtype,subcaption}
\usepackage{color, soul}
\sethlcolor{yellow}

\begin{document}
\onecolumn
\begin{frontmatter}
\journal{Journal}
\title{Large-eddy simulations of wind-driven cross ventilation, Part1: validation and sensitivity study \vspace{-10pt}}

\author[stanford]{Yunjae Hwang \corref{label1}}
\author[stanford]{Catherine Gorl\'e}

\address[stanford]{Wind Engineering Laboratory, Department of Civil and Environmental Engineering, Stanford University, Stanford, CA 94305, USA \vspace{-20pt}}
\cortext[label1]{Corresponding author, yunjaeh@stanford.edu}

\singlespacing
\begin{abstract} 
Natural ventilation is gaining popularity in response to an increasing demand for a sustainable and healthy built environment, but the design of a naturally ventilated building can be challenging due to the inherent variability in the operating conditions that determine the natural ventilation flow. Large-eddy simulations (LES) have significant potential as an analysis method for natural ventilation flow, since they can provide an accurate prediction of turbulent flow at any location in the computational domain. However, the simulations can be computationally expensive, and few validation and sensitivity studies have been reported. The objectives of this study are to validate LES of wind-driven cross-ventilation and to quantify the sensitivity of the solution to the grid resolution and the inflow boundary conditions. We perform LES for an isolated building with two openings, using three different grid resolutions and two different inflow conditions with varying turbulence intensities. Predictions of the ventilation rate are compared to a reference wind-tunnel experiment available from literature, and we also quantify the age of air and ventilation efficiency. The results show that a sufficiently fine grid resolution is needed to provide accurate predictions of the detailed flow pattern and the age of air, while the inflow condition is found to affect the standard deviation of the instantaneous ventilation rate. However, for the cross-ventilation case modeled in this paper, the prediction of the mean ventilation flow rate is very robust, showing negligible sensitivity to the grid resolution or the inflow characteristics.
\end{abstract}

\end{frontmatter}

\onehalfspacing
\section{Introduction} 
Natural ventilation is gaining popularity in response to an increasing demand for a sustainable and healthy built-environment. The use of natural ventilation has significant potential for reducing building energy consumption~\citep{emmerich2001natural, artmann2007climatic, ramponi2014energy}, and it can decrease the risk of respiratory infections such as pneumonia and COVID-19~\citep{ram2014household, urrego2015impact, weaver2017pilot, bhagat2020effects}. The primary objective of natural ventilation is to improve indoor air quality by replacing indoor air with fresh outdoor air, where the airflow is driven by the natural forces of wind and buoyancy. Given the highly variable operating conditions in terms of both the turbulent wind and the temperature field, the prediction of natural ventilation can be challenging~\citep{jiang2002effect, van2010effect, liu2019cfd}. Experimental studies have provided important insight into natural ventilation flow patterns for different configurations, but computational fluid dynamics (CFD) is increasingly popular as an analysis method due to several advantages over experimental methods~\citep{karava2011airflow, ramponi2012cfd, padilla2018assessment}.
First, CFD provides complete access to the flow solution in the entire computational domain, supporting accurate assessment of natural ventilation flow rates, detailed indoor flow patterns, and local ventilation measures such as the age of air. Importantly, these quantities can be obtained without any of the disturbances that may occur during an experiment because of intrusive measurement techniques or reflections or shadows when using optical techniques such as particle image velocimetry (PIV) near ventilation openings~\citep{karava2008thesis}. Second, the advance in computational resources has made it feasible to use CFD for parametric analysis, where the influence of changes in e.g. wind conditions or geometrical configurations can be evaluated using a large number of simulations that are run in parallel.

Ventilation studies have been performed with both Reynolds-averaged Navier-Stokes (RANS) and large-eddy simulation (LES). The RANS approach solves for the Reynolds-averaged field quantities, requiring a model to close the Reynolds stress term. RANS is computationally less expensive than LES, and has been widely employed to assess natural ventilation flow rates and airflow patterns~\citep{shirzadi2019wind, peren2015cfd, peren2015impact, peren2015impact2, peren2016cfd} and to study indoor air quality~\citep{buratti2011mean, tominaga2015wind, tominaga2016wind}. Furthermore, extensive solution verification and sensitivity analyses have been performed~\citep{ramponi2012cfd, ramponi2012cfd2, van2017ransles}. However, the accuracy of RANS models can be significantly reduced when the instantaneous effects of turbulence play an important role in the ventilation process. RANS also has difficulty in accurately predicting the flow field around building geometries, particularly in separation and recirculation regions~\citep{jiang2001study}. 

To overcome the limitations of RANS, several ventilation studies~\citep{jiang2001study, jiang2003buoyancy, jiang2003natural, seifert2006calculation, caciolo2012ransles, van2017ransles} have adopted LES, which solves for the time-dependent turbulent wind. LES solves the filtered governing equations to resolve the larger turbulence scales, while small-scale fluctuations are represented with a subgrid-scale (SGS) model. A few studies directly compared the performance of RANS and LES for ventilation simulations and they all agreed on (1) the superior performance of LES compared to RANS because the averaging process in RANS cancels out instantaneous effects and (2) the increased computational cost of LES (at least one order of magnitude greater than RANS) because it solves unsteady equations and requires a sufficiently fine resolution in space and time~\citep{jiang2001study, jiang2003buoyancy, evola2006computational, caciolo2012ransles, van2017ransles}. 
Albeit the enhanced performance, the computational burden of LES has been a limiting factor to its widespread deployment, and further validation and sensitivity analysis should be performed to investigate the robustness of the predictions for natural ventilation flow. 

In this study, we perform LES for an isolated building with wind-driven cross ventilation, reproducing a reference wind-tunnel measurement conducted by Karava, et al.~\citep{karava2011airflow}. The experiment has previously been employed for validation of CFD models using both RANS and LES. Ramponi, et al.~\citep{ramponi2012cfd, ramponi2012cfd2} used the experimental data for the validation of steady RANS and performed a comprehensive sensitivity analysis to assess the effect the domain size, mesh resolution, turbulence model, and boundary conditions on the solution. Peren, et al.,~\citep{peren2015cfd, peren2015impact, peren2015impact2, peren2016cfd} replicated the experiment in RANS to validate the setup of their simulations, and then used the validated model to study the influence of the shape, height and pitch of the roof on indoor ventilation. Tong, et al.,~\citep{tong2016quantifying, tong2016defining} used the experimental data to validate the setup of their LES prior to using it to define the influence region of indoor ventilation and to study the effect of traffic-related air pollution on the indoor air quality. The impact of LES input parameters such as the inflow boundary conditions has not yet been investigated. 

The objective of the study presented in this paper (Part I) is to further validate LES of wind-driven cross-ventilation, including an investigation of the sensitivity of the solution to the grid resolution and the inflow boundary conditions. The sensitivity analysis considers various measures of ventilation, including time-averaged and instantaneous ventilation flow rates, the local age of air, and the ventilation efficiency. The sensitivity to the grid resolution is investigated by performing simulations with three different meshes. The sensitivity to the inflow conditions considers two inflow conditions that are identical in terms of the mean velocity profile, but have different turbulence intensities in the incoming wind field. In the accompanying paper (Part II), the validated LES will be used to further analyze the performance of a variety of natural ventilation configurations, using the different ventilation metrics. The results for these different configurations will also be interpreted to determine which metrics are more effective for comparing the different ventilation configurations in terms of opening size and position, and wind direction.

The remainder of this paper is organized as follows. Section 2 introduces the reference wind-tunnel experiment for the validation of our CFD model. Section 3 discusses the setup of LES, including the governing equations, computational domain and grid, and boundary and inflow conditions. Section 4 defines the different measures of ventilation used in the current study. In Section 5, we present and analyze the results. Lastly, conclusions and objectives for future research are illustrated in Section 6.

\section{Description of reference experiment}~\label{sec:reference_experiment}
\begin{figure}[hbt!]
\begin{center}
    \includegraphics[width=0.8\textwidth]{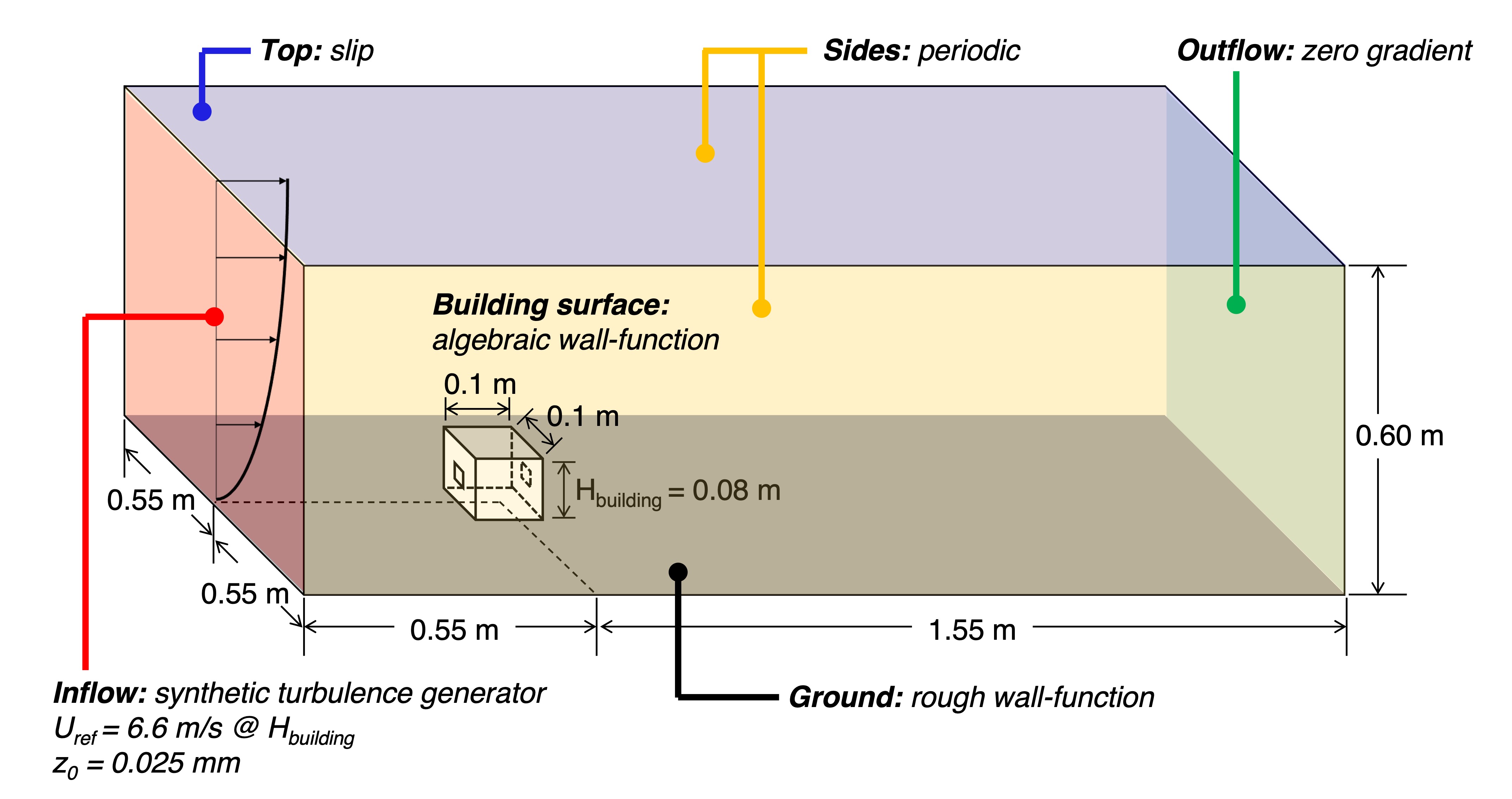}
\end{center}
\caption{Computational domain and boundary conditions for the large-eddy simulations}
\label{fig:domain_bcs}
\end{figure}

The LES setup, shown in Figure~\ref{fig:domain_bcs}, reproduces reference experiments of wind-driven cross ventilation in an isolated building performed in the boundary layer wind tunnel at Concordia University by~\citep{karava2011airflow}.
The model was set-up at the downstream end of the 1.8 m wide by 1.8 m high test section 
The building model had dimensions 100 mm $\times$ 100 mm $\times$ 80 mm and a wall thickness of $t$=2 mm, corresponding to 20 m $\times$ 20 m $\times$ 16 m and $t$=0.4 m in full-scale. A variety of canonical cross ventilation layouts were tested, where each layout had two openings located on opposite or adjacent walls. The combinations of openings tested varied in terms of three parameters. First, different heights along the façade were considered, e.g., bottom, center and top, with the center of the opening at h = 20 mm, 40 mm, and 60 mm, respectively. Second, the wall porosity ($A_{opening}/A_{wall}$) was varied from 2.5\% to 20\% by modifying the opening width for a fixed height of 18 mm. Third, different inlet-to-outlet ratios ($A_{inlet}/A_{outlet}$), ranging from 0.25 to 8 were tested.


The incoming boundary layer was characterized by measuring streamwise velocity and turbulence intensity (TI) profiles in the empty wind tunnel at the intended model location. 
The mean velocity profile has a reference wind speed at building height ($U_{ref}$) equal to 6.6 m/s and a roughness length ($z_0$) equal to 0.025 mm. The streamwise TI is approximately 11\% at the building height and decreases with height. The reference experiment does not provide information regarding the spanwise and vertical TI or regarding the turbulence length or time scales. 

The available measurements include particle image velocimetry (PIV) for the velocity field on both horizontal and vertical planes, hot-film anemometry for the ventilation rate, and pressure measurements at multiple locations on interior wall surfaces. For validation of the simulation results, we use the ventilation rate estimated from both the PIV and the hot-film measurements, and a non-dimensional velocity profile measured along the center-line of the building geometry. It is noted that the PIV measurements may produce inaccurate results near the openings when strong gradients occur and optical effects (e.g. shadows or reflections) influence the quality of measurement. For example, Karava, et al. observed discrepancies in velocities measured at the same location but using different measurement planes (i.e., horizontal and vertical planes)~\citep{karava2008thesis}. 

\section{Large-eddy simulation setup} 
This section presents the setup of the large-eddy simulations (LES). The first subsection describes the governing equations and discretization schemes. Subsequently, we discuss the setup of the computational domain and mesh and we introduce the boundary conditions.

\subsection{Governing equations and discretization method}
\subsubsection{Governing equations}
LES applies a filter to the instantaneous field quantities, i.e., the velocity components ($u_i(x,t)$) and the passive scalar ($C(x,t)$), splitting them into filtered $\widetilde{\langle\cdot\rangle}$ and sub-filtered (sub-grid) components ${\langle\cdot\rangle}'$: $u_i(x,t)=\widetilde{u}_i(x,t)+u_i'(x,t)$; $C(x,t)=\widetilde{C}(x,t)+C'(x,t)$. This procedure results in the following filtered equations for conservation of mass and momentum:
\begin{equation}\label{eq:continuity}
    \frac{\partial \tilde{\rho}}{\partial t} + \frac{\partial \tilde{\rho} \widetilde{u}_i}{\partial x_i}=0,
\end{equation}
\begin{equation}\label{eq:momentum}
    \frac{\partial\tilde{\rho}\widetilde{u}_i}{\partial t}
    +\frac{\partial\tilde{\rho}\widetilde{u}_i\widetilde{u}_j}{\partial x_j}
    =-\frac{\partial\tilde{p}}{\partial x_i}
    +\frac{\partial}{\partial x_j} \left[(\mu+\mu_{sgs}) \left(\frac{\partial\widetilde{u}_i}{\partial x_j}+\frac{\partial\widetilde{u}_j}{\partial x_i} -\frac{2}{3}\delta_{ij}\frac{\widetilde{u}_k}{x_k}\right)\right],
\end{equation}
where $\tilde{\rho}$ is the density, $\mu$ is the kinematic viscosity of air, $\mu_{sgs}$ is the sub-grid scale (SGS) viscosity, representing the effect of the subfilter scales on the resolved motions. $\mu_{sgs}$ is modeled using the Vreman SGS model~\citep{vreman2004eddy} as follows: 
\begin{equation}
    \mu_{sgs}=\tilde{\rho} C_{V} \sqrt{\frac{I_B}{\widetilde{A}_{ij}\widetilde{A}_{ij}}},
\end{equation}
where $C_V$ is the Vreman coefficient, $\widetilde{A}_{ij}=\frac{\partial \widetilde{u}_j}{\partial x_i}$ is the filtered velocity gradient tensor, and $I_B$ is the second invariant of the tensor $\widetilde{B}_{ij}=\Delta^2\widetilde{A}_{ij}\widetilde{A}_{ij}$. 

In addition to the equations for conservation of mass and momentum, a scalar transport equation is solved to assess the local age of air:
\begin{equation}\label{eq:scalar}
    \frac{\partial\tilde{\rho}\widetilde{C}}{\partial t}
    +\frac{\tilde{\rho}\partial\widetilde{u}_j \widetilde{C}}{\partial x_j}
    =\frac{\partial}{\partial x_j} 
    \left[\left(\tilde{\rho}\widetilde{D}+\frac{\mu_{sgs}}{Sc_{sgs}}\right)\frac{\partial \widetilde{C}}{\partial x_j}\right],
\end{equation}
where $\widetilde{D}$ is the molecular diffusion coefficient for the scalar and $\frac{\mu_{sgs}}{Sc_{sgs}}$ represents a subgrid turbulent diffusion coefficient, with $Sc_{sgs}$ the subgrid turbulent Schmidt number, which is set equal to 1.0.

\subsubsection{Discretization method and solution procedure}
The governing equations are solved using the CharLES Helmholtz solver developed by Cascade Technologies, Inc~\citep{charles}. The solver adopts a second-order central discretization in space and a second-order implicit time-advancement with a fixed time-step size. The time-step size is such that the maximum CFL number is always lower than 1.0. The statistics of the quantities of interest are estimated using the flow solution obtained over 250 $\tau_{ref}$, after an initial burn-in period of at least 100 $\tau_{ref}$, where $\tau_{ref}$ is the flow-through time for the target house (the ratio of the width of the house to the wind speed at the reference height, i.e., $D_{House}/U_{ref}=0.1/6.6\approx 0.015$ sec).

\subsection{Computational domain and mesh}
\subsubsection{Computational domain}
The dimensions of the CFD domain are based on the COST action 732 best practice guidelines~\citep{franke2007cost} to avoid any unintended effects of the boundary conditions on the flow solution. The resulting domain has dimensions $W \times D \times H =$ 1.1 m $\times$ 2.1 m $\times$ 0.6 m, which is equal to $13.75 H_{building} \times 26.25 H_{building} \times 7.5H_{building}$, where $H_{building}=$ 0.08 m is the height of the building geometry.
The inflow boundary is located at a distance of 6.875 $H_{building}$ from the center of the house, while the outflow boundary is located 18.375 $H_{building}$ downstream from the same location. The two lateral boundaries are at least 6$H_{building}$ away from the building. These dimensions satisfy the recommendations in the COST 732 guideline for any orientation of the building geometry, thereby supporting simulations for all wind directions that will be considered in this study.


\subsubsection{Computational mesh and grid dependency study}
The computational grid is generated with the CharLES mesh generator and a grid sensitivity study is conducted to determine the influence of the mesh resolution on the results. 
The sensitivity study employs three different meshes, i.e., coarse, baseline and fine. These meshes only differ in the background cell size; all other settings such as the location and size and of the refinement box and the number of transition layers between refinement levels are identical. The local refinement box around the building is sufficiently large to encompass the upstream standing vortex, the flow separations on the sides and the top of the building, and the wake region downstream. 
Table \ref{tab:grid_sensitivity_study_summary} summarizes the background and smallest cell sizes, as well as the total number of cells for the three cases used in the grid sensitivity study. The baseline and fine meshes meet the requirement of having at least 10 cells across an area of interest, i.e., along the edges of the openings~\citep{franke2007cost}, but the coarse mesh has only five to six cells along the opening height. 
\begin{table}[htb!]
    \caption{Summary of grid sensitivity study: background cell size, smallest cell size and the number of cells in the unit of million control volumes}
    \label{tab:grid_sensitivity_study_summary}    
    \begin{center}
    \begin{tabular}{|c|c|c|c|}
    \hline 
Cases& Coarse & Baseline & Fine  \\ \hline 
Background cell size [mm] & 32  & 16  & 8   \\ \hline 
Smallest cell size [mm] & 3.0 & 1.5 & 0.75  \\ \hline 
Number of cells [M cells] & 0.483 & 3.21 & 24.1 \\ \hline
    \end{tabular}
    \end{center}
\end{table}

\subsection{Boundary conditions}
\subsubsection{Wall, side and top boundary conditions}
Figure~\ref{fig:domain_bcs} visualizes the setup of the LES for the sensitivity analysis and the validation study, including the boundary conditions. The ground and building surfaces are no-slip walls, and wall functions are used to calculated the friction velocity. A smooth wall algebraic wall model is used on all building surfaces, while a rough wall function for a neutral atmospheric boundary layer (ABL) with a roughness length ($z_0$) of 0.025 mm is specified at the ground boundary. The two lateral boundaries are periodic and a slip condition is applied on the top boundary. The outlet boundary condition is set to a zero gradient condition. 

\subsubsection{Inflow conditions and target profiles} \label{sec:inflow_conditions}
The inflow boundary condition is designed to reproduce the ABL velocity generated in the wind tunnel experiment (the target profiles). The boundary condition combines a divergence-free digital-filter method proposed by~\cite{xie2008efficient} and~\cite{kim2013divergence} with a gradient based optimization to obtain the target turbulence characteristics at the building location in the domain~\citep{lamberti2018optimizing}. In this section, we first present the target profiles that should be obtained at the building location. Subsequently the optimization technique and its results are presented. 

The digital filter method generates turbulent inflow conditions using inputs for the mean velocity profile, the Reynolds stresses, and the turbulence length- or time-scales. The reference experiment reports the mean velocity and streamwise turbulence intensity (TI) profiles. Information on the spanwise and vertical TIs or the the turbulence length- or time-scales is not available and  is therefore estimated from similarity relationships. 

For the mean velocity profile, a logarithmic velocity profile is fitted to the measurements:
\begin{equation}
    U(z)=\frac{u_*}{\kappa} \log (\frac{z+z_0}{z_0}),
\end{equation}
with the reference velocity of 6.6 m/s at the reference height of 0.08 m and the roughness length of 0.025 mm. The streamwise component of TI is taken from the experimental data, while the spanwise and vertical components can be approximated based on similarity relationships~\citep{stull2012abl}: 
\begin{equation}
    \overline{u'u'}(z) = (TI_u(z) \cdot U(z) )^2, \qquad \overline{v'v'}(z) = \overline{w'w'}(z) = \overline{u'u'}/\sqrt{2}. 
\end{equation}

To estimate the turbulence length scales, we use data available from measurements performed when the wind tunnel was constructed~\citep{stathopoulos1984windtunnel}. The longitudinal length scale ($L_u$) was estimated to be 112 m in full-scale at one sixth of the boundary layer depth. At model scale, this length scale corresponds to 0.28 m at a height of 0.1 m, which is approximately the building height. The spanwise and vertical length scales are estimated using their ratio to the streamwise component:
\begin{equation}
    L_v=0.2L_u, \qquad L_w=0.3L_u.
\end{equation}
These length scales are converted to turbulent time scales using Taylor's hypothesis:
\begin{equation}
    T_u =L_u \times U_{ref}, \quad T_v = L_v \times U_{ref}, \quad T_w =L_w \times U_{ref}.
\end{equation}

\subsubsection{Inflow optimization}\label{subsubsec:inflow_optimization}
A limitation of the inflow generation method is that the artificially generated turbulence does not correspond to a solution of the governing equations. Therefore, the turbulence intensities tend to decay along the streamwise direction in the computational domain, often resulting in turbulence characteristics at the area of interest, i.e., at the building location, that are considerably lower than those specified at the inflow. To compensate for this decay, we utilize a gradient-based optimization technique that specifies optimized inflow profiles of turbulence intensity and time scales at the inlet, such that the desired turbulence statistics are retrieved at the target building location~\citep{lamberti2018optimizing}. 

\begin{figure}[htp!]
\begin{center}
    \includegraphics[width=0.9\textwidth]{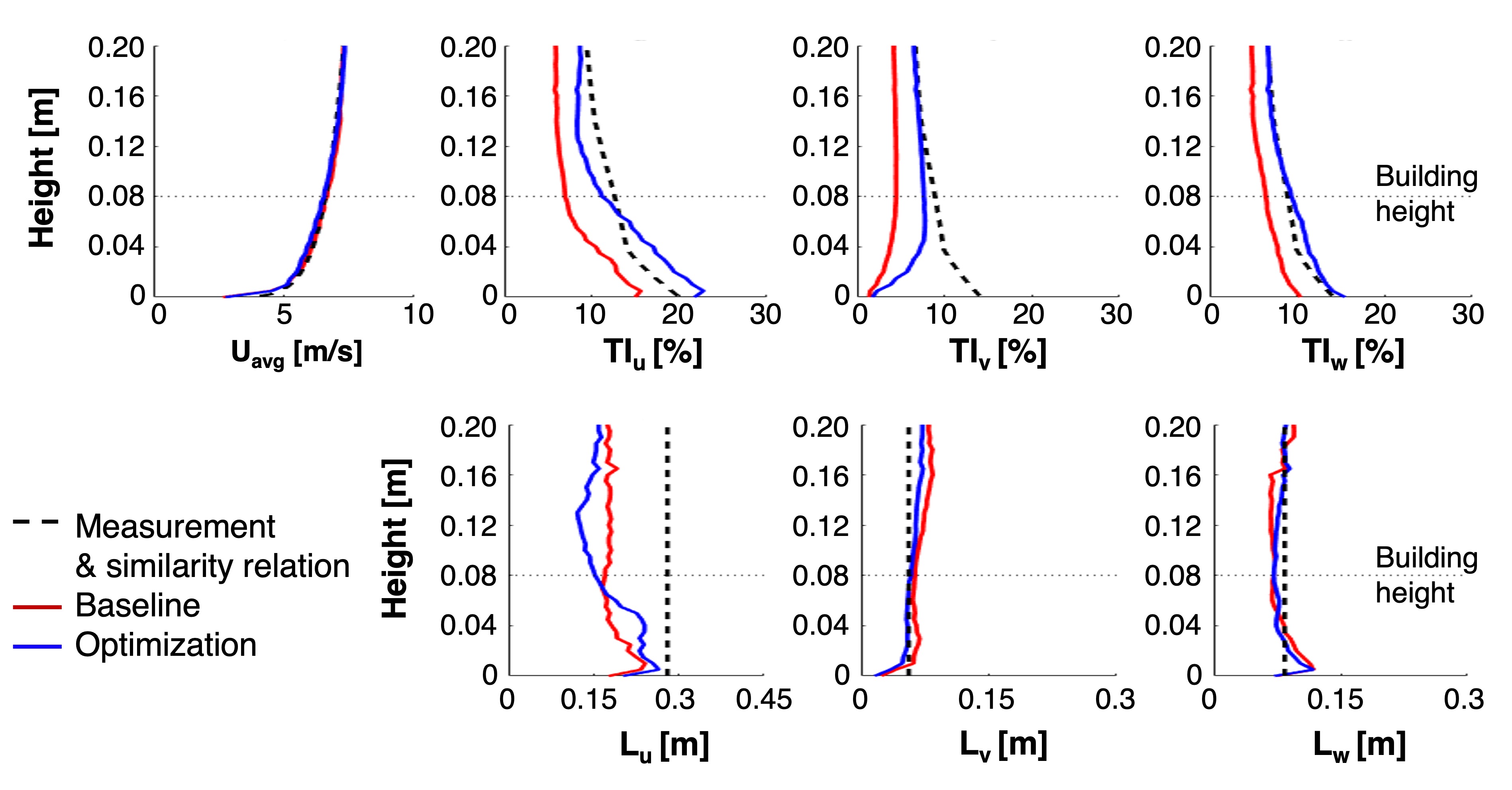}
\end{center}
\caption{Mean velocity, turbulence intensity, and length scale profiles: target profiles (black dashed line), profiles obtained at the building location using the target profiles at the inflow (red solid line), profiles obtained at the building location using the optimized inflow profiles (blue solid line)}
    \label{fig:inflow_optimization}
\end{figure}

Figure~\ref{fig:inflow_optimization} presents the target profiles for the velocity, turbulence intensity and time scales (black dashed lines), and the profiles for the same quantities at the building location (red and blue solid lines). When the target profiles are imposed at the inlet, they decay as the flow advances along the domain and lower turbulence characteristics are achieved at the building location (red line). Combined with the optimization technique, the inflow generator imposes a higher turbulence intensity at the inlet to compensate for the decay. The resulting profiles at the building location (blue line) compare well to the target profiles. The inflow sensitivity analysis (Section~\ref{subsec:result_inflow_sensitivity}) will compare the LES predictions obtained with the baseline and the optimized inflow conditions.

\section{Quantity of interest: measures of ventilation} \label{sec:quantity_of_interest}
\subsection{Instantaneous and time-averaged ventilation rates} 
The primary objective of ventilation is to replace indoor air with fresh outdoor air. Hence, the net amount of the indoor-outdoor air exchange is an important ventilation measure. This subsection outlines different strategies for calculating the air exchange rate, using either the pressure difference across the openings or integration of the normal velocity through the openings.

Natural ventilation is driven by pressure differences across openings. For a single room with two identical opening and steady flow conditions, the ventilation rate can be obtained from the analytical solution of an envelope flow model: 
\begin{equation} \label{eq:Q_p_avg}
   Q_{p, avg}=C_d \cdot A \cdot \sqrt{\frac{2\cdot|P_1-P_2|}{\rho}}, 
\end{equation} 
where $Q_{p,avg}$ is a time-averaged ventilation rate calculated using the time-averaged pressure difference across the openings $|P_1-P_2|$, with $P_{1,2} =\frac{1}{T} \int_0^T \int_{A} p_{1,2}(t)\,dA\,dt$. $C_d$ is the still-air discharge coefficient of the openings, $\rho$ is the density of air and $A$ is the area of the openings. This relationship has been widely adopted in ventilation studies~\citep{seifert2006calculation, karava2006impact, evola2006computational, karava2011airflow}, but it has two limitations. First, its accuracy can be compromised by uncertainty in $C_d$. The coefficient is commonly assumed to be a constant of 0.61, corresponding to the still-air discharge coefficient of a fully open window. In reality, the value can be affected by multiple parameters, including the outdoor wind direction~\citep{karava2004wind}. 
Second, the equation is based on a steady-state flow assumption, and it does not account for ventilation due to turbulent air exchange. 

To eliminate the parametric uncertainty related to $C_d$ in Eq.~\ref{eq:Q_p_avg}, the time-averaged flow rate through the openings can be calculated by integrating the time-averaged normal velocity across the openings:  
\begin{equation} \label{eq:Q_u_avg}
    Q_{u,avg} = \frac{1}{2} (\int\limits_{A_1} |\overline{\mathbf{u}} \cdot \mathbf{n}_{1}|\,d\text{A} + \int\limits_{A_2} |\overline{\mathbf{u}} \cdot \mathbf{n}_{2}|\,d\text{A}),
\end{equation}
where $\overline{\mathbf{u}}$ and $\mathbf{n}$ are the time-averaged velocity vector and the area normal vector, respectively. The equation calculates the time-averaged ventilation flow rate as half of the sum of the net volume flow rate through the two openings. 
This calculation is commonly adopted in both experimental approaches and CFD simulations using RANS, which generally provide the time-averaged velocity fields (e.g.~\cite{karava2011airflow, caciolo2012ransles, van2017ransles}). While this formulation removes the uncertainty due to $C_d$ in Eq.~\ref{eq:Q_p_avg}, it still assumes that the ventilation is driven by a uni-directional mean flow. Hence, Eq.~\ref{eq:Q_u_avg} can still lead to inaccurate predictions of the ventilation rate when turbulence plays an important role, e.g. in single-sided ventilation scenarios~\citep{jiang2001study}. 

To overcome the limitation related to time-averaging the velocity field through the openings, one can directly estimate the instantaneous ventilation rate from integration of the instantaneous velocity field $\mathbf{u}(t)$ across the openings:
\begin{equation}\label{eq:Q_u_ins}
    Q_{u,ins}(t) = \frac{1}{2} (\int\limits_{A_1} |\mathbf{u}(t) \cdot \mathbf{n}_{1}| \,d\text{A}_1 + \int\limits_{A_2} |\mathbf{u}(t) \cdot \mathbf{n}_{2}| \,d\text{A}_2).
\end{equation} 
From the time series of the instantaneous ventilation rate, the statistics, i.e., mean and standard deviation, can be estimated. Given a consistent, steady, flow direction over time, as is the case in the cross-ventilation configuration considered in this Part I paper, the time-average of $Q_{u,ins}(t)$ is equal to $Q_{u,avg}$. Hence, in this paper we will only present the results based on $Q_{u,ins}(t)$. However, in certain configurations, when the flow direction is variable, the time-average of $Q_{u,ins}(t)$ can be significantly different from $Q_{u,avg}$, since averaging the velocity fields in Eq.~\ref{eq:Q_u_avg} filters out turbulent fluctuations that contribute to air exchange. ~\cite{jiang2001study} demonstrated that these differences are particularly important for single-sided ventilation. In the accompanying Part 2 paper, we will further investigate these discrepancies in cross-ventilation configurations for which turbulent air exchange plays an important role.

The ventilation rate is widely used to assess natural ventilation performance, but one limitation is that it does not provide any insight into internal air motion. In some cases, e.g. when re-circulation zones are presents in the indoor space, a high ventilation rate does not imply that the entire space is well-ventilated and the air quality can be locally reduced. The following sections introduce quantities that give additional insight into the spatial distribution of the ventilation pattern.

\subsection{Age of air}~\label{subsec:age_of_air}
The age of air can be used to assess spatial variability in the rate at which indoor air is replaced, providing a local measure for indoor air quality. The local age of air 
is defined as the average time that an air parcel at a certain location has been in the indoor space. 

To estimate the local age of air, we adopt a tracer concentration decay technique, which is straightforward to implement in the CFD simulations. The method first initializes a tracer (the passive scalar $\widetilde{C}$ in Eq.~\ref{eq:scalar}) with a non-zero concentration $C_0$ in the indoor space. Subsequently, the time evolution of the tracer concentration is recorded, until it returns to the background concentration, which is 0 in the simulations. The age of air at a position $x$ is then calculated from the area under the normalized tracer decay curve:
\begin{equation} \label{eq:age_of_air}
    \tau(x) = \int_0^\infty \frac{\widetilde{C}(x,t)}{C_0} dt,   
\end{equation}
where $\widetilde{C}(x,t)$ is the tracer concentration at a point $x$ at time $t$. The age of air calculation is performed once a quasi steady-state solution for the flow-field has been obtained, i.e. after the initial burn-in period.

\begin{figure}[htp!]
    \centering
    \includegraphics[width=\textwidth]{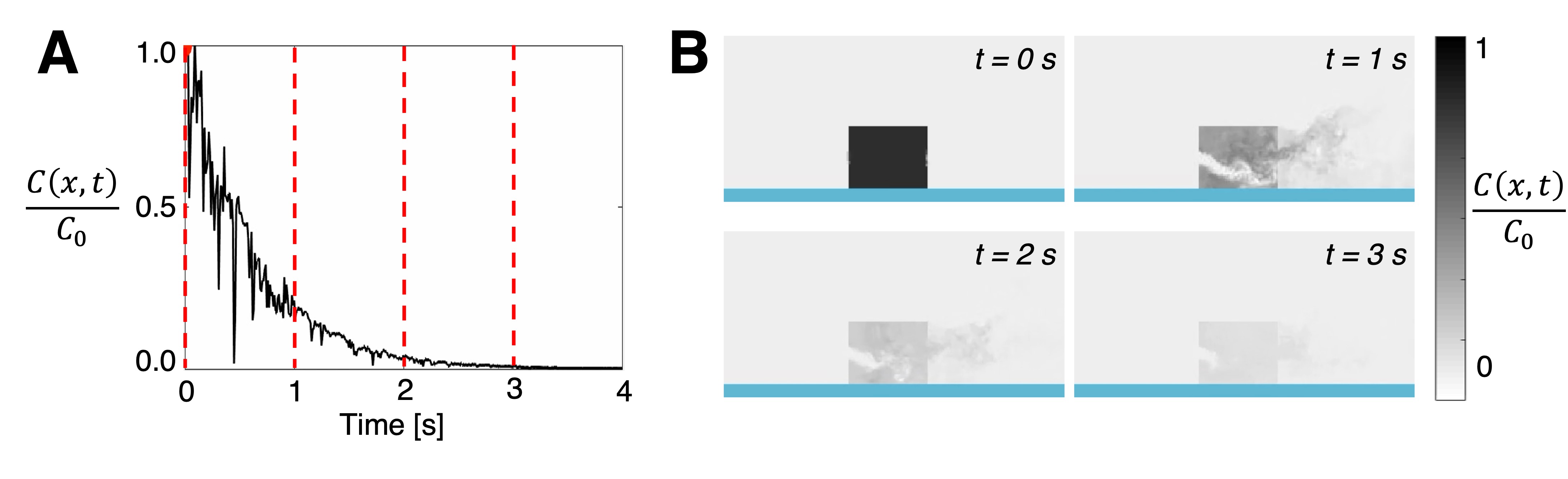}
    \caption{Graphical representation of age of air calculation: \textbf{(A)} Concentration decay curve for passive scalar at the center of house; and \textbf{(B)} indoor concentration at four time steps}
    \label{fig:age_of_air}
\end{figure}
Figure \ref{fig:age_of_air} graphically presents the calculation process of the age of air, showing (A) the time-evolution of the normalized concentration at one indoor point (the center of the building), and (B) the contours of the tracer concentration on the vertical plane crossing the house at four different points in time. The contour plots demonstrate that the indoor space is unequally ventilated, such that the age of air varies by location: $\tau$ is zero near the inlet where the air flushes out very quickly, while the value is higher than average in poorly ventilated areas such as the recirculation regions near the floor and the roof. 
The example in Fig.~\ref{fig:age_of_air} demonstrates how the age of air enables us to identify well and poorly ventilated locations in the space, unlike the ventilation rate, which provides a single value for the entire space. 

\subsection{Ventilation efficiency}\label{subsec:ventilation_efficiency}
Ventilation efficiency aims to quantify how effectively ventilation occurs within a space. This efficiency can be defined in different ways~\citep{sandberg1981ventilation, murakami1992new}, but a common approach is to define it as follows:
\begin{equation} 
    \epsilon = \frac{\langle \tau \rangle_{min}}{\langle \tau \rangle},
\end{equation}
where $\langle \tau \rangle_{min}$ is the minimum value of the spatial average of the age of air that can be hypothetically be achieved with a certain flow rate, and $\langle \tau \rangle$ is the actual spatial average of the age of air. To specify the hypothetical minimum $\langle \tau \rangle_{min}$, we consider the case where all particles at the inlet at time $t=0$ reach the outlet in the minimum possible residence time $\tau_{r,min}$. This corresponds to a piston flow, with the minimum residence time equal to the nominal time $\tau_n = \frac{V}{Q}$. Using the relationship that the spatial average of the age of air $\langle \tau \rangle_{min}$ is equal to $\tau_{r,min}/2 = \tau_n/2$, we obtain:
\begin{equation} \label{eq:vent_efficiency}
\epsilon = \frac{\tau_n}{2 \langle \tau \rangle}.
\end{equation}

The ventilation efficiency provides a measure for comparing the performance of different ventilation solutions in terms of the overall 'freshness' of the air for a given ventilation flow rate. 
For the piston flow, $\epsilon=1$ and the entire space is ventilated at the ideal rate, such that the time taken to replace all the air within the volume is equal to half $\langle \tau_n \rangle $. For a perfectly mixed flow, it is assumed that once air enters the space it is immediately uniformly distributed throughout the space. In this case, the spatial average of the age of air, $\langle \tau \rangle$, is equal to the nominal time, $\tau_n$, such that $\epsilon=0.5$. 


\section{Results}\label{sec:results}
This section presents the results of two sensitivity analyses: sensitivity to the grid resolution, and sensitivity to the inflow conditions. For both analyses, a qualitative comparison of the difference in the predicted flow patterns and a quantitative comparison of the different ventilation metrics is performed. 

\subsection{Grid sensitivity study}\label{subsec:result_grid_sensitivity}
\subsubsection{Influence of grid resolution on flow field}\label{subsubsec:result_grid_flow_field}
\begin{figure}[hbp!]
\begin{center}
\includegraphics[width=\textwidth]{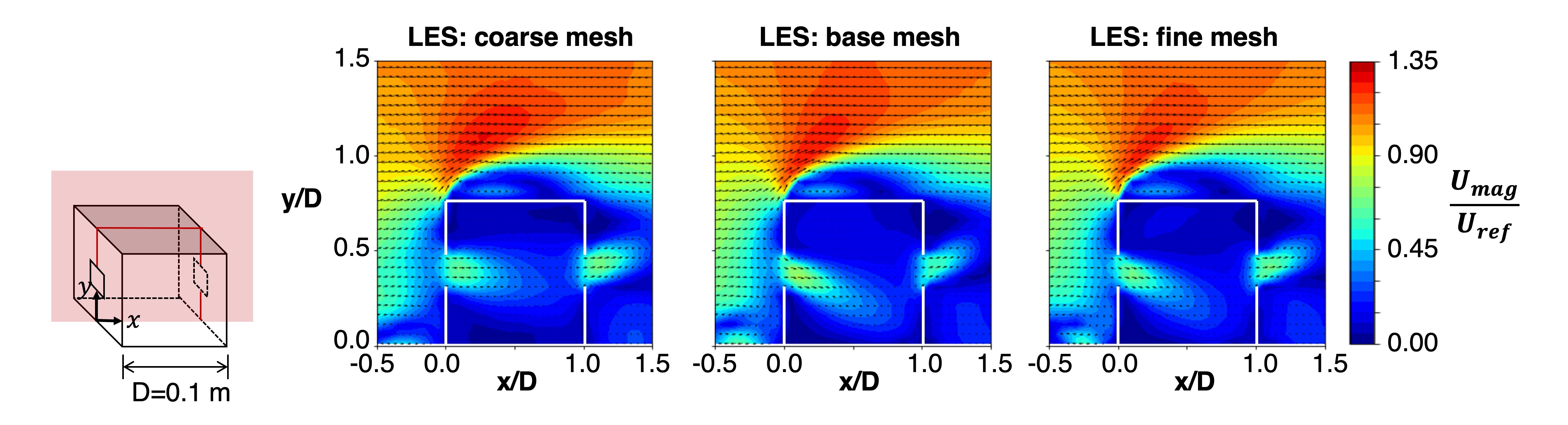}
\end{center}
\caption{Velocity fields obtained from simulations using the coarse (left), base (center) and fine (right) meshes. Each plot displays contours of velocity magnitude contours together with a quiver plot to visualize the flow direction} 
\label{fig:result_grid_flow_field}
\end{figure}
The influence of the mesh resolution on the LES results was determined by performing simulations with gradually refined computational grids, using the optimized inflow conditions shown in Figure~\ref{fig:inflow_optimization}. Figure~\ref{fig:result_grid_flow_field} presents the resulting velocity fields on a vertical plane through the center of the building geometry, showing contours of velocity magnitude and quiver plots to indicate the flow direction. The presentation of the results focuses on this vertical plane, since it depicts important flow features around the building geometry and shows the differences between the three simulations most clearly. 
Qualitatively, the results obtained from the base and fine meshes compare well, while the coarse mesh results show more noticeable discrepancies. All three results show some common, typical flow features around the house, including the flow separation and reattachment on the roof. However, the coarse mesh provides a slightly different prediction of the standing vortex upstream of the windward facade and of the flow just downstream of the inlet window. Overall this indicates that the limited resolution in the coarse mesh, with a small number of cells (5-6) along the height of the windows can reduce the accuracy of the flow prediction. This result confirms the suggested minimum requirement of 10 cells across an opening~\citep{franke2007cost}. 

\begin{figure}[ht!]
    \centering
    \includegraphics[width=0.7\textwidth]{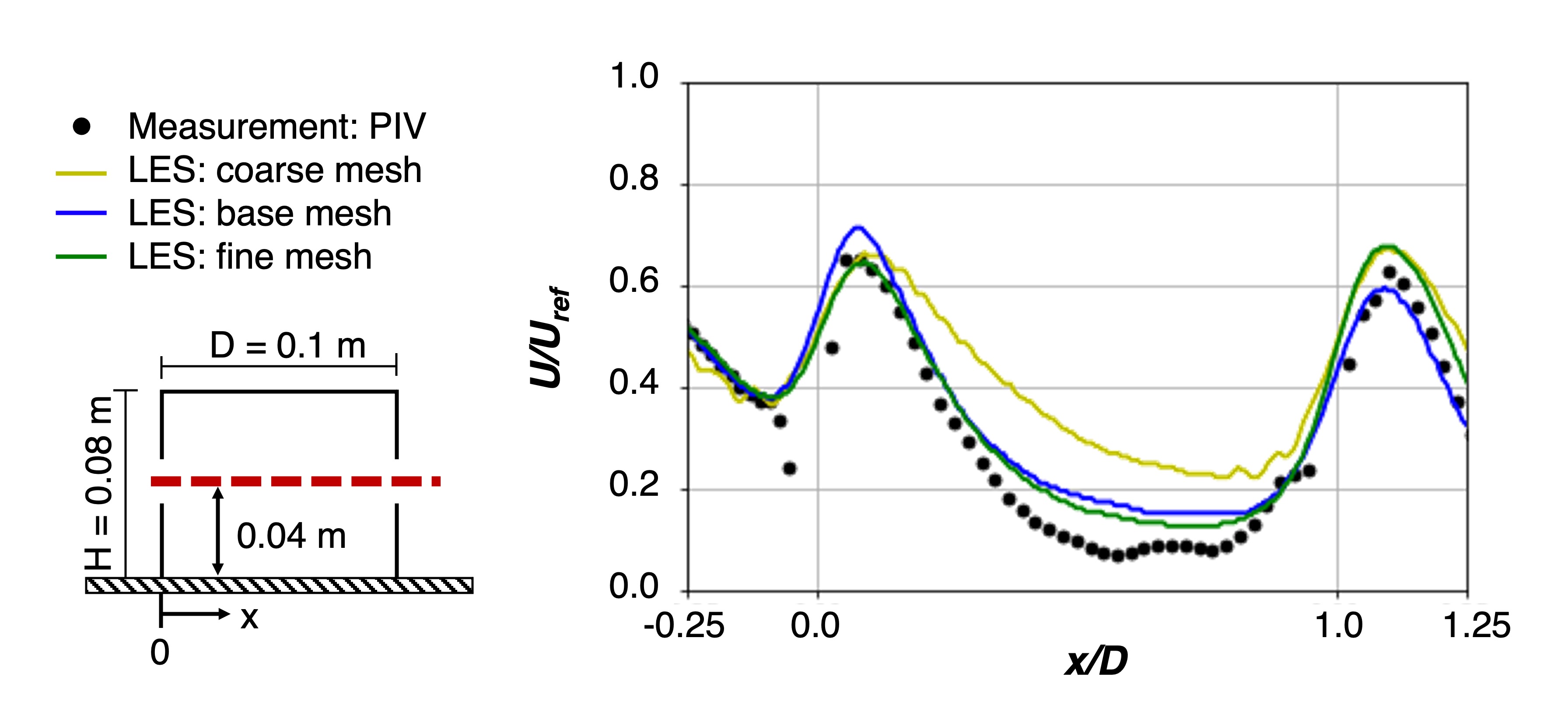}
    \caption{Non-dimensional velocity profile along the center line predicted using the coarse, base and fine meshes}
    \label{fig:result_grid_velocity_profile}
\end{figure}

A more quantitative comparison of the influence of the grid resolution on the velocity field is shown in Figure~\ref{fig:result_grid_velocity_profile}. The plots shows a velocity profile along a horizontal line through the center of the window opening. In addition to the three simulation results it includes the PIV measurements by~\citep{karava2011airflow}. Similarly to the velocity contours, the plots indicate close agreement between the results obtained with the base and fine meshes, while there are more significant differences with the coarse mesh results. The base and fine mesh results also compare better to the PIV data. The maximum discrepancies between the LES and PIV occur in the lowest velocity region and are limited to about 0.1$U/U_{ref}$. Focusing on the maximum velocities that occur just downstream of the openings along this horizontal line, the results do indicate a difference on the order of ~0.1$U/U_{ref}$ between the base and fine meshes. The following section determines whether this difference in the predicted local velocity also results in differences in the predicted ventilation measures. 

\subsubsection{Influence of grid resolution on ventilation measures}~\label{subsubsec:result_grid_vent_measures}

This section first evaluates the influence of the grid resolution on the ventilation rates calculated using Eqs.~\ref{eq:Q_p_avg} and~\ref{eq:Q_u_ins}. Subsequently the effect on the age of air and the ventilation efficiency is quantified.
\begin{figure}[ht!]
\begin{center}
\includegraphics[width=1.0\textwidth]{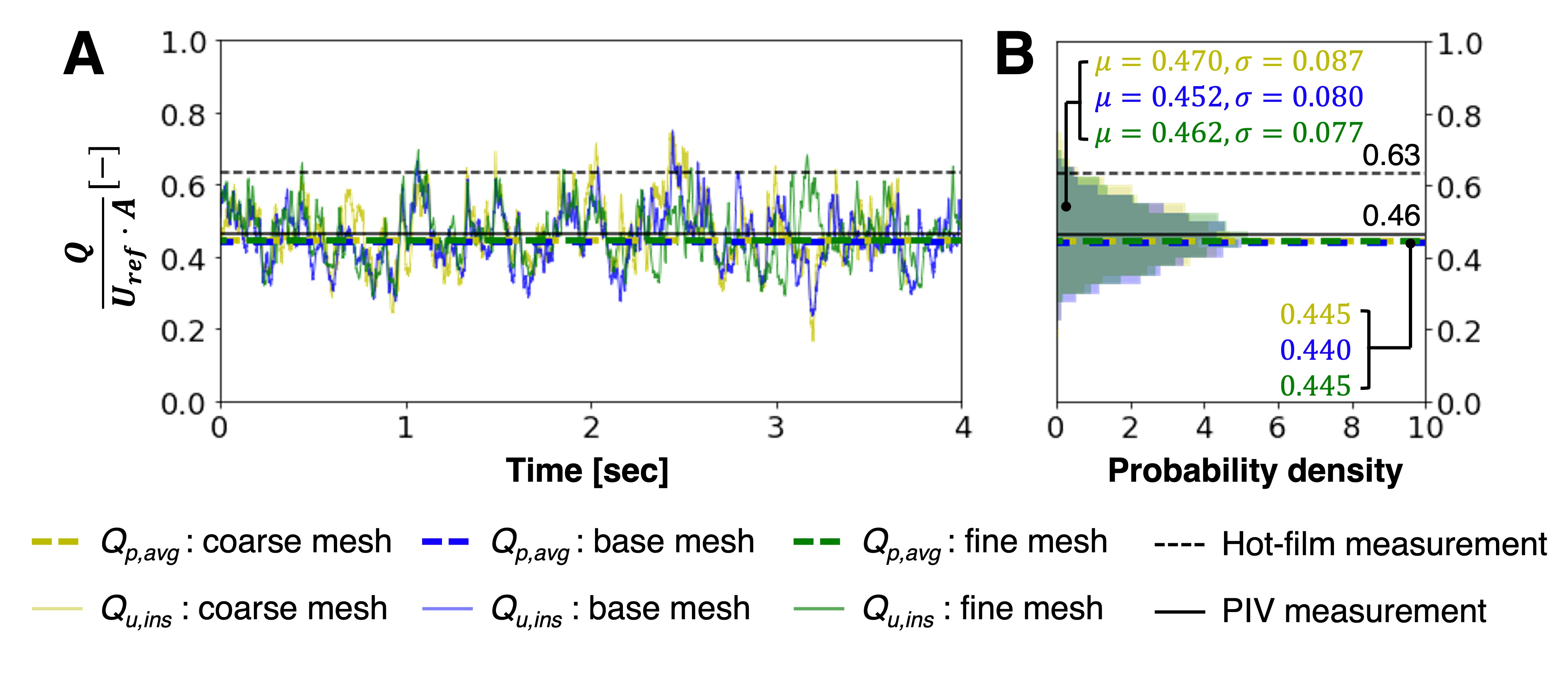}
\end{center}
\caption{Grid sensitivity study results for the dimensional ventilation rates: \textbf{(A)} Time series of ventilation rate calculated from the velocity field, together with average ventilation rate estimated from the time-averaged pressure field, and the experimental values; \textbf{(B)} Corresponding distributions and mean values.}
\label{fig:result_grid_vent_rate}
\end{figure}
Figure~\ref{fig:result_grid_vent_rate} presents the results for the dimensionless ventilation rates $Q/(U_{ref}\cdot A)$ obtained using the different grid resolutions, including a comparison to the ventilation rates obtained from the PIV and hot-film measurements in the experiment. Figure~\ref{fig:result_grid_vent_rate}A displays time-series of $Q_{u,ins}(t)/(U_{ref}\cdot A)$, i.e. the dimensionless ventilation rates calculated using the instantaneous velocity. In addition, the time-averaged dimensionless ventilation rates $Q_{p,avg}/(U_{ref}\cdot A)$ calculated using the time-averaged pressure differences are shown, together with the values obtained from the experiments. Figure~\ref{fig:result_grid_vent_rate}B shows the same information, but the time series are depicted as distributions, and their mean values and standard deviations are reported. The results obtained using the different LESs and the PIV measurements are in close (~3\%) agreement. The hot-film measurement predicts a higher ventilation rate, which could be because the turbulent three-dimensional complex flow field near the ventilation openings introduces significant uncertainty in the hot-film-based ventilation rate estimate~\citep{karava2008thesis}.
Comparing the mean values of $Q_{u,ins}(t)/(U_{ref}\cdot A)$ for the coarse, base, and fine meshes, the values are 0.470, 0.452 and 0.462, respectively; the corresponding values for $Q_{p,avg}/(U_{ref}\cdot A)$ are 0.445, 0.440 and 0.445. The maximum difference between the different meshes is ~3\%, indicating limited dependency of the predicted ventilation rates on the grid resolution. In fact, the ventilation rates are more sensitive to the calculation method, i.e. using the velocity vs the pressure, than to the grid resolution. The difference between both methods can be attributed to the use of the still-air discharge coefficient value $C_d$=0.61 when computing $Q_{p,avg}$ with equation~\ref{eq:Q_p_avg}; the actual discharge coefficient is influenced by the specific flow pattern near the windows. In the current configuration the differences remain limited to ~6\%, but larger discrepancies can occur for different configuration (e.g. different wind directions). The dependency of $C_d$ on the ventilation configuration will be further investigated in the accompanying Part II paper. 

\begin{figure}[htp!]
    \centering
    \includegraphics[width=1.0\textwidth]{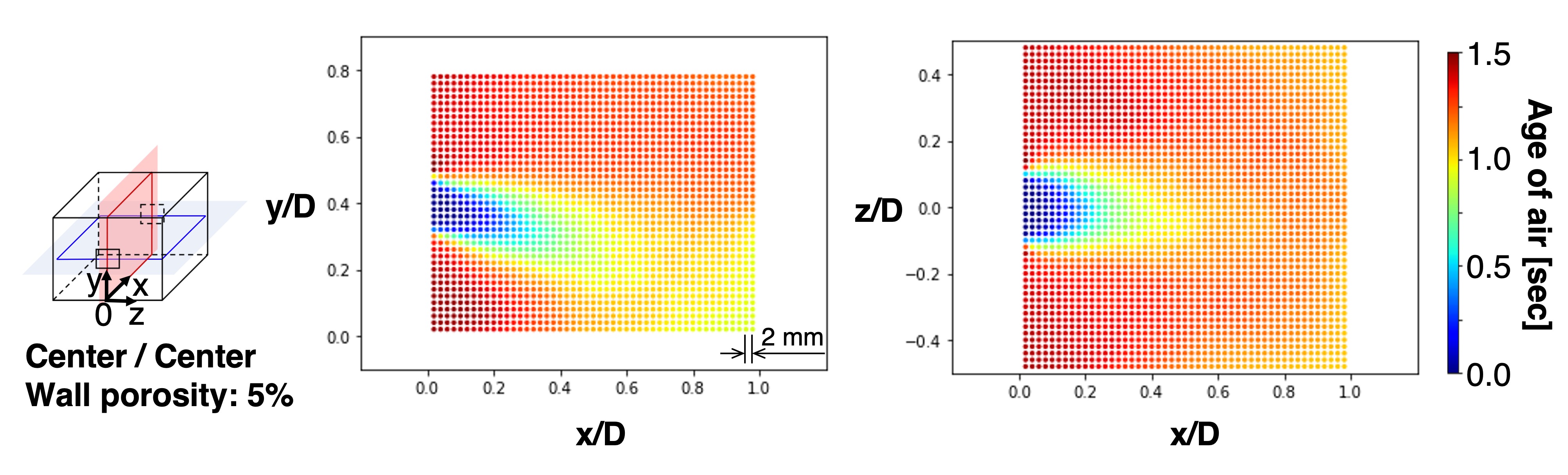}
    \caption{Distribution of mean age of air at sample points on vertical and horizontal planes through the center of the building.}
    \label{fig:result_aoa_2d_planes}
\end{figure}
\begin{figure}[ht!]
\begin{center}
\includegraphics[width=0.6\textwidth]{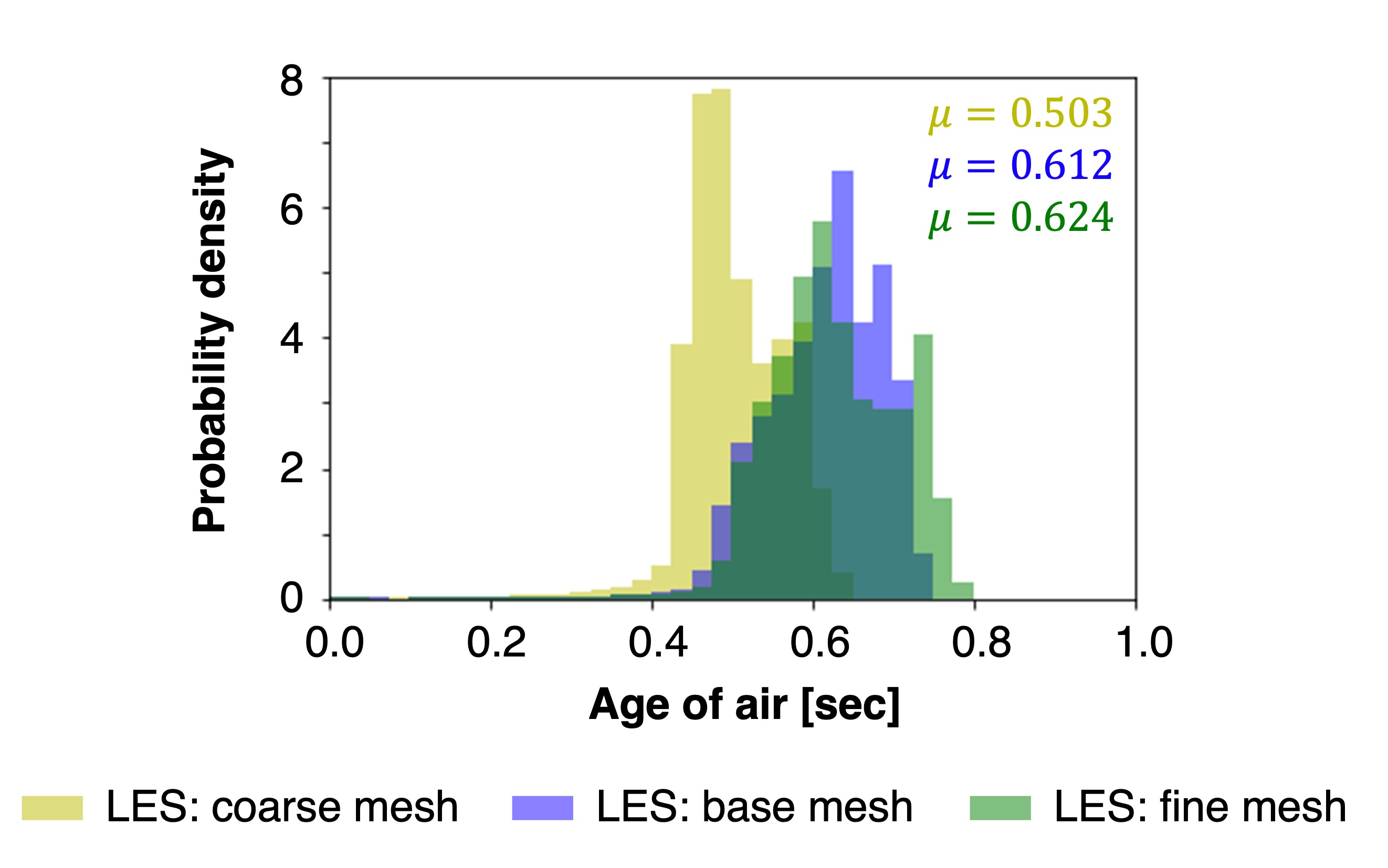}
\end{center}
\caption{Frequency distribution of age of air using the coarse, base, and fine meshes.}   \label{fig:result_grid_aoa}
\end{figure}

The age of air is a function of space and it is generally visualized using a frequency distribution, which represents the overall ventilation status of the space. To provide some insight into the spatial distribution of the age of air, Figure~\ref{fig:result_aoa_2d_planes} first displays its value at specific sample points on a horizontal and a vertical plane, both through the center of the building, taken from the simulation using the base mesh. The distribution of the age of air on the vertical plane looks similar to the velocity field on this plane (Fig.~\ref{fig:result_grid_flow_field}), since generally the mean velocity and the age of air are inversely correlated, i.e. locations with a higher velocity magnitude are more likely to be well-ventilated and have a low age of air.
Figure~\ref{fig:result_grid_aoa} displays the frequency distributions of the age of air collected at approximately 94,000 uniformly distributed points in the building for all three LES results. The distribution of the age of air obtained from the coarse simulation looks significantly different from the results obtained with the other two meshes. The mean value of the age of air with the coarse mesh is estimated to be 0.503, while the base and fine meshes predict 0.612 and 0.624, respectively. This results indicates that compared to the overall ventilation rate, the age of air is significantly more sensitive to the grid resolution, with a ~20\% difference in the mean values obtained from the coarse and fine mesh. 

To conclude this section, Table~\ref{tab:result_grid_measures} summarizes the results for the different ventilation metrics: $Q_{p,avg}$, the mean and standard deviation of $Q_{u,ins}(t)$ and the spatial average of the age of air ($\langle \tau \rangle$). The table also includes the corresponding ventilation efficiency ($\epsilon$). Similar to the comparison of the velocity field in Section~\ref{subsubsec:result_grid_flow_field}, the base and fine meshes yield very similar results with a maximum error of 4\% for the ventilation efficiency. The prediction accuracy of the coarse mesh varies depending on the ventilation measure. The mean ventilation rates $Q_{p,avg}$ and $Q_{u,ins}$ are well predicted; however, quantities that depend on local predictions of the concentration decay have higher differences. The ~20\% underprediction of the mean age of air results in an overprediction of the ventilation efficiency by ~22\% compared to the fine mesh. 
\begin{table}[htbp!]
\caption{Summary of ventilation measures obtained using the coarse, medium, and fine meshes}\label{tab:result_grid_measures}
\begin{center}
    \begin{tabular}{|l|c|c|c|c|} \hline
                        & Coarse            & Base              & Fine              \\ \hline 
$Q_{p,avg}$ [m$^3$/s]   & 2.852       & 2.860       & 2.908       \\ \hline 
$Q_{u,ins}(t)$ [m$^3$/s]   & $\mu$=3.104       & $\mu$=2.982       & $\mu$=3.052       \\ 
                    & $\sigma$=0.574    & $\sigma$=0.526    & $\sigma$=0.506    \\ \hline
$\langle \tau \rangle$ [s] & 0.503      & 0.612             & 0.624  \\ \hline
$\epsilon$ [-]      & 0.619             & 0.530             & 0.507  \\ \hline
\end{tabular}
\end{center}    
\end{table}

The grid sensitivity analysis presented in this section indicates that if the ventilation rate is the primary quantity of interest, the coarse simulation can be sufficient, given its accuracy in predicting both $Q_{p,avg}$ and $Q_{u,ins}(t)$. However, if an accurate estimate of the age of air and ventilation efficiency is required, the base mesh configuration provides the best choice in terms of the balance between computational cost and accuracy.

\subsection{Inflow sensitivity study}~\label{subsec:result_inflow_sensitivity}
\subsubsection{Influence of inflow condition on flow field}~\label{subsec:result_inflow_flow_field}

\begin{figure}[ht!]
\begin{center}
\includegraphics[width=1.0\textwidth]{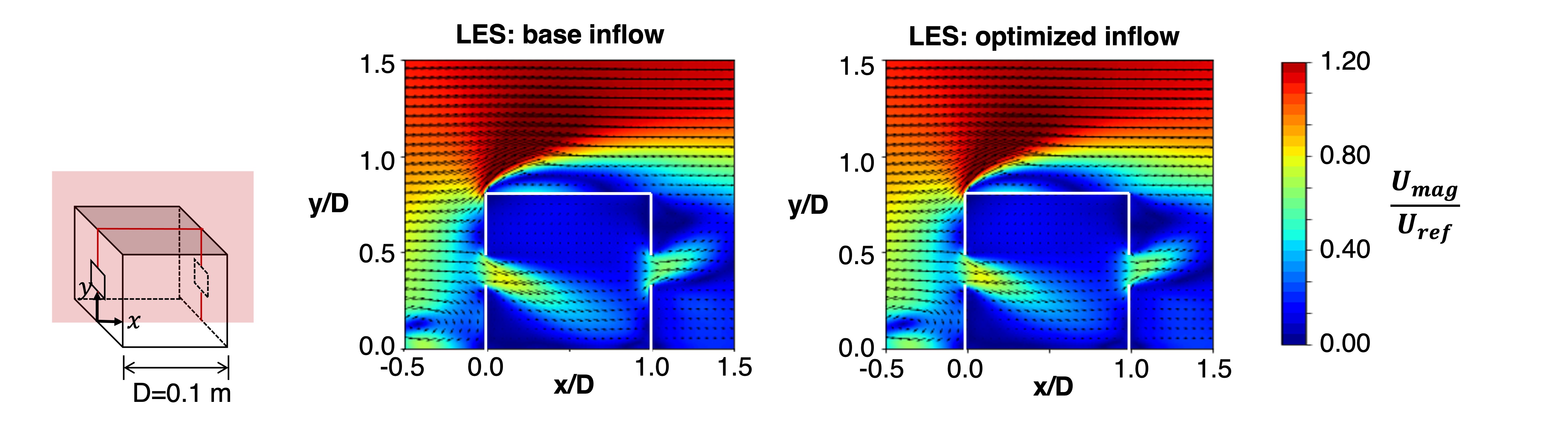}
\end{center}
\caption{Velocity magnitude contour with quiver plot for flow direction using the base inflow (left) and optimized inflow (right).} 
\label{fig:flow_pattern_inflow_sensitivity}
\end{figure}

The influence of the inflow boundary conditions on the LES results was determined by performing simulations with the baseline and optimized inflow conditions shown in Fig.~\ref{fig:inflow_optimization}. Figure~\ref{fig:flow_pattern_inflow_sensitivity} presents the resulting velocity fields on the vertical plane through the center of the building geometry, showing contours of velocity magnitude and quiver plots to indicate the flow direction. The two velocity patterns look very similar in terms of all the typical flow features, including the standing vortex in front of the building, flow separation and reattachment on the roof, and the ventilation inflow and outflow direction. 

\begin{figure}[htbp!]
\begin{center}
\includegraphics[width=0.7\textwidth]{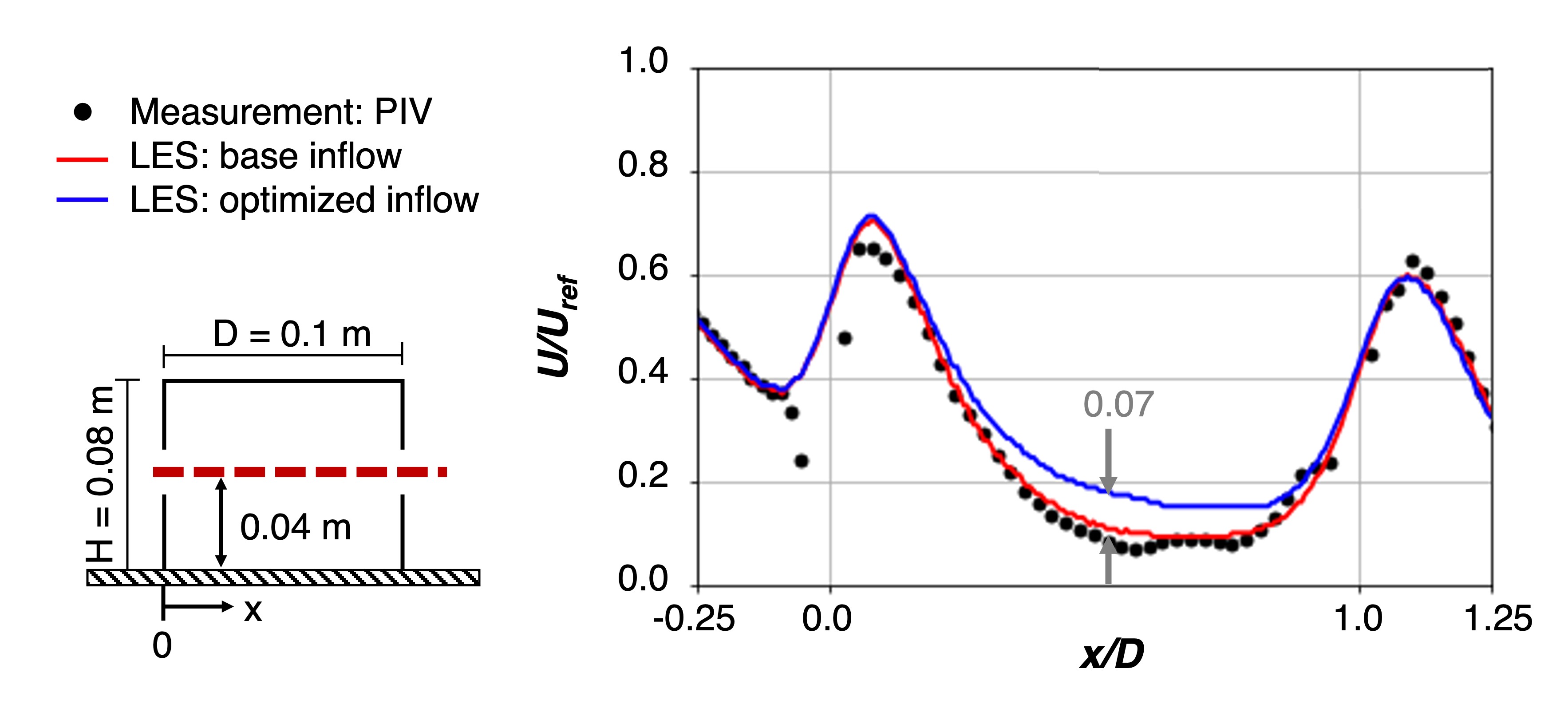}
\end{center}
\caption{Non-dimensional velocity profile along the center line predicted using the base and optimized inflow conditions}
\label{fig:velocity_profile_inflow_sensitivity}
\end{figure}

Figure~\ref{fig:velocity_profile_inflow_sensitivity} shows a more quantitative comparison of the influence of the inflow on the velocity field, plotting the velocity profile along a horizontal line through the center of the window opening. In addition to the three simulation results it includes the PIV measurements by~\citep{karava2011airflow}. The influence of the inflow boundary condition is confirmed to be small. Differences only occur in the low velocity regions of the flow, and they are less than 0.1$U/U_{ref}$. This limited effect of the inflow boundary condition on the mean velocity field can be attributed to the fact that only the intensity, and to a limited extent the length scales, of the turbulent fluctuations are changed, i.e. the mean inflow velocity profiles are the same. This finding is in line with a similar sensitivity analysis for the prediction of the pressure coefficients on a building, where it was found that the turbulent wind statistics do not affect the mean pressure coefficient, but they do affect the pressure coefficient fluctuations~\citep{lamberti2020sensitivity}.

\subsubsection{Influence of inflow condition on ventilation measures}~\label{subsec:result_inflow_vent_measures}

\begin{figure}[htb!]
\begin{center}
\includegraphics[width=1.0\textwidth]{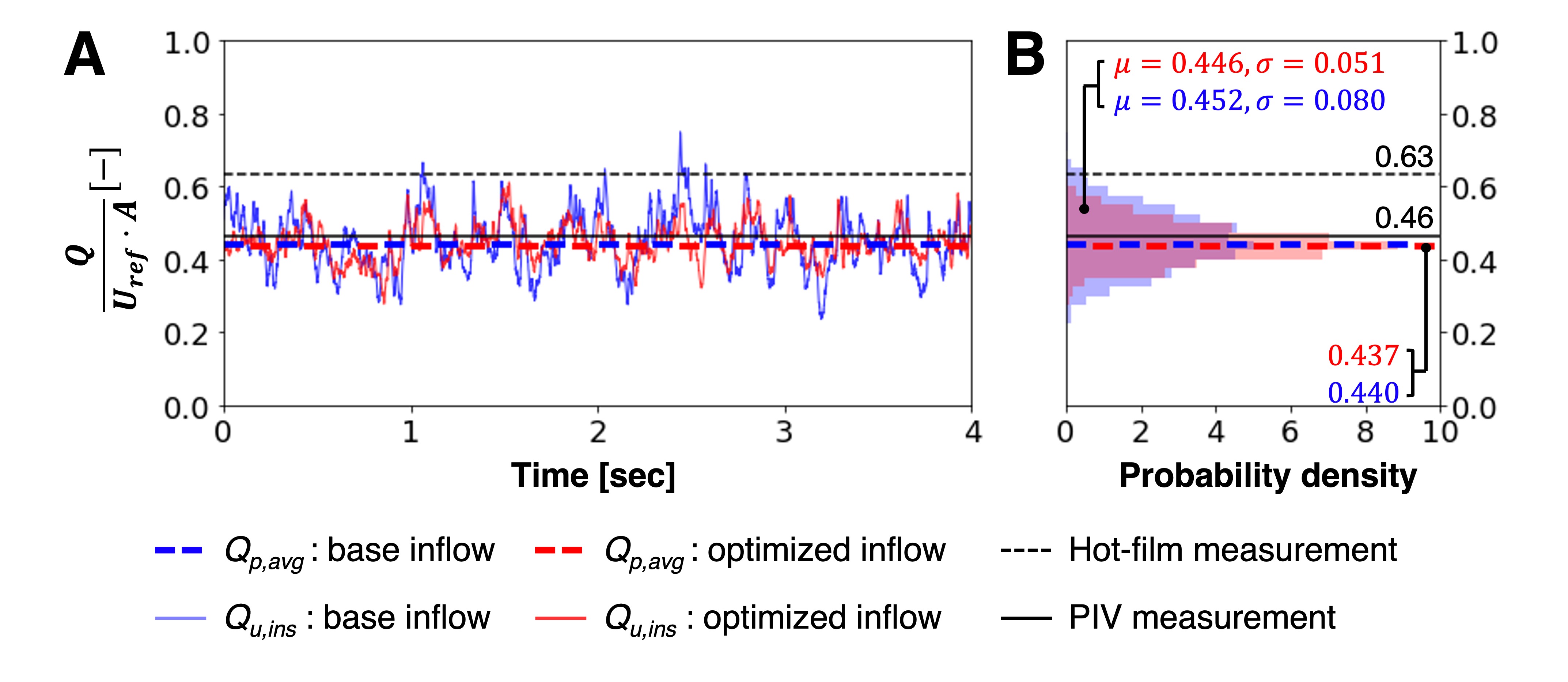}
\end{center}
\caption{Inflow sensitivity study results for the dimensional ventilation rates: \textbf{(A)} Time series of ventilation rate calculated from the velocity field, together with average ventilation rate estimated from the time-averaged pressure field, and the experimental values; \textbf{(B)} Corresponding distributions and mean values.} 
\label{fig:Q_inflow_sensitivity}
\end{figure}

This section first evaluates the influence of the inflow conditions on the ventilation rates calculated using Eqs.~\ref{eq:Q_p_avg} and~\ref{eq:Q_u_ins}. Subsequently the effect on the age of air and the ventilation efficiency is quantified. Figure~\ref{fig:Q_inflow_sensitivity} presents the results for the dimensionless ventilation rates obtained using the two inflow conditions, including a comparison to the values obtained from the PIV and hot-film measurements in the experiment. Figure~\ref{fig:result_grid_vent_rate}A displays time-series of $Q_{u,ins}(t)/(U_{ref}\cdot A)$, i.e. the dimensionless ventilation rates calculated using the instantaneous velocity. In addition, the time-averaged dimensionless ventilation rates $Q_{p,avg}/(U_{ref}\cdot A)$ calculated using the time-averaged pressure differences are shown, together with the values obtained from the experiments. Figure~\ref{fig:result_grid_vent_rate}B shows the same information, but the time series are depicted as distributions, and their mean values and standard deviations are reported. The results for the mean ventilation rates obtained using the different LESs are in close agreement and compare well to the PIV measurement, with less then 5\% discrepancy. 
While the mean ventilation rate predictions are not affected by the change in the inflow boundary condition, the standard deviation of $Q_{u,ins}(t)$ does change considerably: the optimized inflow condition predicts a standard deviation of 0.8, while the base inflow condition predicts a standard deviation of 0.51. Clearly, the increased turbulence intensity in the optimized inflow results in increased turbulent fluctuations in the flow rate through the windows. While this change in the turbulence fluctuations does not affect the time-average ventilation rate in this specific cross-ventilation configuration, it might play a more dominant role when the instantaneous fluctuations are a dominant contribution to the overall ventilation rate. For example, if one would consider a wind direction of 90$^\circ$ for the present configuration, a higher turbulence intensity in the inflow could affect the average ventilation rate more significantly. The importance of the contribution of turbulence fluctuations to the average ventilation rate will be further investigated in the Part II paper, where the ventilation flow rate estimated with time-averaged quantities ($Q_{p,avg}$ and $Q_{u,avg}$) will be compared to the temporal mean of the instantaneous ventilation rate ($Q_{u,ins}(t)$) for different wind directions.

\begin{figure}[htb!]
\begin{center}
    \includegraphics[width=0.6\textwidth]{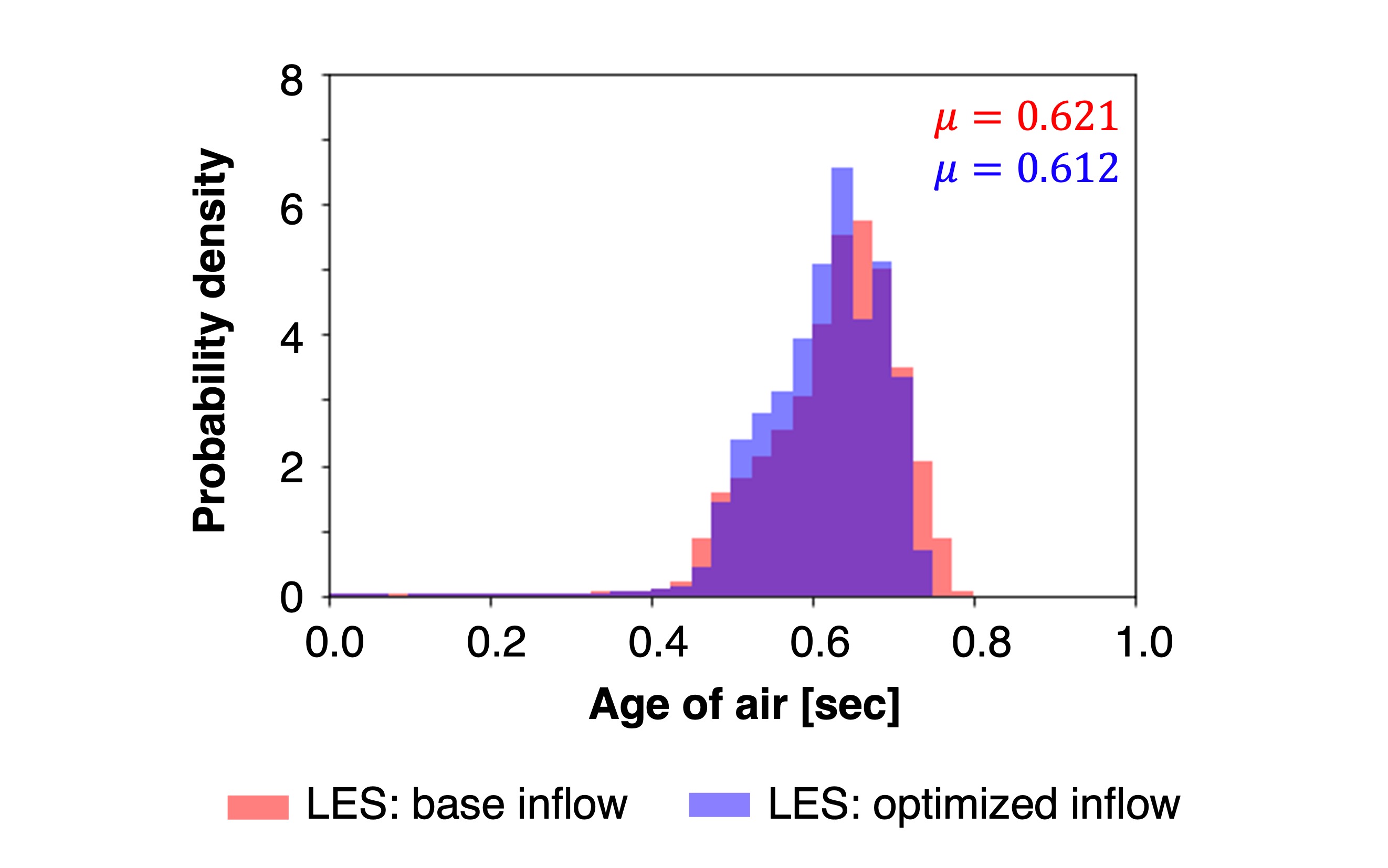}
\end{center}    
    \caption{Frequency distribution of age of air using two different inflow conditions}   
    \label{fig:result_inflow_aoa}
\end{figure}
Figure~\ref{fig:result_inflow_aoa} presents the effect of the inflow conditions on the age of air, where the distributions are constructed from the age values at uniformly distributed points in the building. The age of air does not seem sensitive to the inflow conditions, with the difference between the mean values only around 1.5\%. 

Lastly, Table~\ref{tab:result_inflow_measures} summarizes the results for the different ventilation metrics, including the ventilation efficiency. Comparison to the results of the grid sensitivity analysis in Table~\ref{tab:result_grid_measures} indicates that the average ventilation measures are less sensitive to the inflow turbulence than to the grid resolution, with a maximum discrepancy of 1.5\% for the spatial average of the age of air. However, the standard deviation of $Q_{u,ins}(t)$ is significantly more sensitive to the inflow changes than to the grid resolution. This finding indicates that for configurations where turbulent fluctuations provide a significant contribution to the overall ventilation rate, the turbulence characteristic of the incoming wind could have a strong impact on the resulting ventilation rate.

\begin{table}[htb!]
\caption{Summary of ventilation measures for the inflow sensitivity study} \label{tab:result_inflow_measures}    
\begin{center}
    \begin{tabular}{|c|c|c|} \hline
                    &  Baseline inflow  & Optimized inflow     \\ \hline
$Q_{p,avg}$         & 2.864             & 2.860                      \\ \hline
$Q_{u,ins}(t)$   & $\mu$= 2.944      & $\mu$= 2.982                      \\ 
                    & $\sigma$= 0.338   & $\sigma$= 0.526                   \\ \hline
$\langle \tau_n \rangle$ & 0.621        & 0.612                             \\ \hline
$ \epsilon$         & 0.529             & 0.530                             \\ \hline
    \end{tabular}
\end{center}
\end{table}

\section{Conclusion and future work}

We have investigated the sensitivity of large-eddy simulations (LESs) for wind-driven cross-ventilation in an isolated building to the grid resolution and the inflow boundary conditions. The LESs reproduce a wind-tunnel experiment available in a literature, supporting validation of the results. The grid sensitivity analysis indicates that the simulations are quite robust in terms of predicting the average ventilation rates, with all results providing an accurate prediction of the ventilation rate measured by PIV in the experiments. However, more local quantities, such as the local velocity downstream of the window opening and the local age of air, do exhibit some grid-dependency, with the coarse mesh underpredicting the spatial average of the age of air by ~20\%. For the solver and numerical schemes used in this paper, we found that this grid sensitivity becomes negligibly small when having at least 10 cells in each direction along the window openings. 

The inflow sensitivity analysis focused on the effect of the turbulence intensity in the incoming wind on the results. It was shown that measures for the time-average ventilation rate exhibit little dependency on this turbulence intensity, with a maximum discrepancy of 1.5\% for the spatial average of the age of air. However, the standard deviation of the instantaneous ventilation rate is significantly more sensitive to the inflow changes, increasing by ~20\% for the inflow conditions with the increased turbulence intensity. This finding indicates that for configurations where turbulent fluctuations provide a significant contribution to the overall ventilation rate, the turbulence characteristic of the incoming wind could have a strong impact on the resulting ventilation rate. Our study shows that it is important to accurately reproduce the turbulence in the atmospheric wind when pursuing validation of LES against experiments. To support careful validation, experiments should ideally report accurate data on the inflow turbulence characteristics, including all three turbulence intensities and time scales. 

The results presented in this paper indicate the significant potential of LES for assessing natural ventilation performance. The information that can be obtained in terms of instantaneous ventilation rates and local age of air is very difficult to obtain from wind tunnel experiments, and the baseline grid resolution provides accurate predictions at a reasonable computational cost. In Part II of this paper, we will leverage this validated simulation setup to quantify the performance of different natural ventilation configurations using the average and local measures of ventilation introduced in this paper.

\section*{Acknowledgments}
This research was funded by a seed grant from the Stanford Woods Institute Environmental Venture Projects program and supported by the Stanford Center at the Incheon Global Campus (SCIGC) funded by the Ministry of Trade, Industry, and Energy of the Republic of Korea and managed by the Incheon Free Economic Zone Authority.
The authors are thankful for the high-performance computing support from Stampede2 provided by Texas Advanced Computing Center (TACC).

\bibliographystyle{Frontiers-Harvard} 
\bibliography{reference}

\end{document}